\documentclass{osa-article}
\usepackage{subcaption}
\usepackage[font=normalsize,labelfont={bf}]{caption}
\usepackage{hyperref}
\usepackage{float}
\usepackage{xcolor} 

\journal{oe}

\begin{document}

\title{Femtosecond Temporal Phase-Resolved Nonlinear Optical Spectroscopy in Molecules with Lock-in Enabled Phase Tracking}

\author{Siddhant Pandey,\authormark{1} Francis Walz,\authormark{1} and Niranjan Shivaram \authormark{1,2,*}}

\address{\authormark{1} Department of Physics and Astronomy, Purdue University, West Lafayette, Indiana 47907, USA\\
\authormark{2} Purdue Quantum Science and Engineering Institute, Purdue University, West Lafayette, Indiana 47907, USA}

\email{\authormark{*}niranjan@purdue.edu} 


\begin{abstract}
We describe an experiment to measure the emitted real-time electric field from an ultrafast third-order nonlinear optical interaction in molecules, using a phase-tracked spectral interferometry scheme. By combining a software lock-in amplification based spectrometer with spectral interferometry, we measure the electric field of the nonlinear optical signal from rotationally excited gas-phase molecules. The lock-in spectrometer allows selective measurement of signals of interest with improved signal-to-noise ratio, while rejecting any unwanted incoherent background. The nonlinear optical signal interferes with a known reference pulse on the spectrometer, which allows measurement of ultraweak signal electric fields. Further, we show that lock-in detection enables correction of slow interferometric drifts by utilizing a multidimensional measurement space. Thus, interferometric stability is achieved without the need for active stabilization, which typically utilizes an independent phase drift measurement. We present data from an experiment in impulsively aligned molecules that demonstrates the important features of our scheme. The scheme can be applied to study ultrafast dynamics in laser excited systems in the gas, liquid, and solid phases of matter.
\end{abstract}

\section{Introduction}
The availability of intense ultrafast laser pulses has made it possible to investigate ultrafast nonlinear optical phenomena in atomic, molecular, solid, and biological systems. The nonlinear optical response of a material results from the coherent addition of multiple excitation pathways \cite{boyd2008}. Multidimensional spectroscopy is often used to help disentangle the dynamics resulting from different pathways \cite{oliver2018}. In addition, signals from different pathways typically have different propagation directions and can be further isolated spatially. However, many such pulse-averaged measurements integrate the information contained in the amplitude and phase of the emitted signal electric field (E-field). Measurement of the complete E-field  of the signal from a light-matter interaction has been demonstrated to offer novel insight into the interaction \cite{sommer2016}. In an ultrafast nonlinear optical process, any emitted light contains spectral and temporal signatures of the interaction. For example, in four-wave mixing (FWM), the emitted signal is related to the induced third-order electronic polarization in the medium and is determined by the third-order nonlinear response tensor $\chi^{(3)}_{ijkl}$. A measurement of the amplitude and phase of such emitted signal fields can be expected to give access to the maximum information that can be gained in an all-optical measurement. Recently, it has been demonstrated that the measurement of the full E-field of the emitted field in a four-wave mixing (FWM) experiment can be used to distinguish between contributions from electronic and nuclear degrees of freedom \cite{walz2022}. Furthermore, molecular frame measurements of the temporal phase of four-wave mixing signals performed at room temperature have been shown to be sensitive to differences in electronic symmetries of different molecules \cite{pandey2024}.

Ultrafast E-field-resolved measurements have been previously demonstrated using attosecond streaking \cite{itatani2002,goulielmakis2004,eckle2008} and direct field sampling \cite{liu2021,liu2022,bionta2021,park2018,hui2022,schiffrin2013,paasch2014,paasch2016,sederberg2020} techniques. Measuring the temporal electric field (E-field) of femto-joule (fJ) level signals is challenging for these methods although measurement of sub-fJ pulses is possible with electro-optic sampling \cite{srivastava2023}. In this work, we report the development of a lock-in enabled spectral interferometry scheme that can measure the ultraweak emitted E-field from an ultrafast nonlinear optical interaction, and demonstrate it in laser aligned gas-phase molecules. Lock-in enabled detection efficiently rejects signals from un-aligned molecules, allowing us to measure very weak signals from aligned molecules.

In the following sections, we start by introducing spectral interferometry for measuring E-fields. We then discuss our implementation of software-based lock-in enabled imaging, which is used to build a lock-in spectrometer. Finally, using experimental measurements of FWM signals from rotationally excited gas-phase molecules, we demonstrate the salient features of using lock-in enabled spectral interferometry for probing ultrafast dynamics in molecules. We conclude by discussing future improvements and scope of this technique.

\section{\label{sec:interferometry}Spectral Interferometry}
Nonlinear optical signals can be detected using a square law detector, like a CCD/CMOS sensor, that only measures the intensity of the incident light. Techniques like optical Kerr effect (OKE) spectroscopy \cite{wieman1976,lotshaw1987,palese1994} typically measure the pulse-time-integrated  real and imaginary parts of the signal, using optical heterodyning (OHD) with a local oscillator (LO). In our experiments we use spectral interferometry for the measurement of very weak (picojoule - femtojoule) femtosecond pulses \cite{fittinghoff1996,chen1997}, using a known external reference pulse. In this measurement, the signal field is combined with the known reference pulse on a spectrometer and the spectral interference fringes are recorded. If the two beams are separated by a time $\tau$ the expression for the measured spectrum is:

\begin{equation}\label{eq:RS}
    S(\omega) = S_{R}(\omega) + S_{S}(\omega) + \sqrt{S_{S}(\omega) S_{R}(\omega)}\cos(\varphi_{SR}(\omega)+\omega \,\tau)
\end{equation}

The subscripts $S$ and $R$ stand for signal and reference, respectively. For convenience, the terms that do not oscillate with time delay $\tau$ in our expressions will be referred to as "DC", and the fast oscillating terms as "AC". $S(\omega)$ is the spectral intensity, $\omega$ is the angular frequency, and $\varphi(\omega)$ is the spectral phase. $\varphi_{SR}$ is the difference between the spectral phases of the signal and reference. To extract the signal phase $\varphi_{S}(\omega)$, we first Fourier transform $S(\omega)$ with respect to $\omega$. The first two terms on the right-hand side of equation \ref{eq:RS} only have a DC component. The last term contains the fringes between the two pulses and has fast oscillations, due to the $\omega\tau$ term in the argument of the cosine. The Fourier transform of $S(\omega)$ has a distinct non-zero frequency peak corresponding to the fringe period. We filter and shift this peak to 0 Hz to remove the $\omega\tau$ term, and then inverse Fourier transform the resulting spectrum \cite{takeda1982} from which we can extract $\varphi_{SR} = \varphi_{S}(\omega) - \varphi_{R}(\omega)$. The reference pulse is intense enough to be fully characterized using a Frequency Resolved Optical Gating (FROG) \cite{trebino1993}. Using the known reference phase $\varphi_{R}(\omega)$ we can extract $\varphi_{S}(\omega)$.

In measurements where a coherent background co-propagates with the signal, equation \ref{eq:RS} needs to be modified. With the addition of a coherent background, equation \ref{eq:RS} becomes,

\begin{equation}\label{eq:RSB}
    \begin{split}
        S(\omega) = \ &S_{S}(\omega) + S_{R}(\omega) + 
        \sqrt{S_{S}(\omega) S_{R}(\omega)}\cos(\varphi_{SR}(\omega)+\omega\, \tau) \\ 
        &+S_{B}(\omega)+\sqrt{S_{B}(\omega) S_{R}(\omega)}\cos(\varphi_{BR}(\omega) + \omega\, \tau) \\
        &+\sqrt{S_{S}(\omega) S_{B}(\omega)}\cos(\varphi_{SB}(\omega))
    \end{split}
\end{equation}

where the subscript $B$ is used to denote the coherent background. $\varphi_{BR}(\omega)$ and $\varphi_{SB}(\omega)$ are the differences in spectral phase between background and reference, and signal and background, respectively. Presence of the coherent background and the additional interference fringes (additional terms in equation \ref{eq:RSB}) it produces on the detector complicate the extraction of the signal spectrum and phase. Although coherent background from the excitation pulses themselves can be spatially removed by using a non-collinear geometry, such separation is not possible between ground- and excited-state signals, which usually obey the same phase-matching conditions \cite{pandey2024}.

\section{\label{sec:lockin_interferometry}Lock-in Enabled Interferometry}

\begin{figure*}
    \centering
    \includegraphics[width=\linewidth]{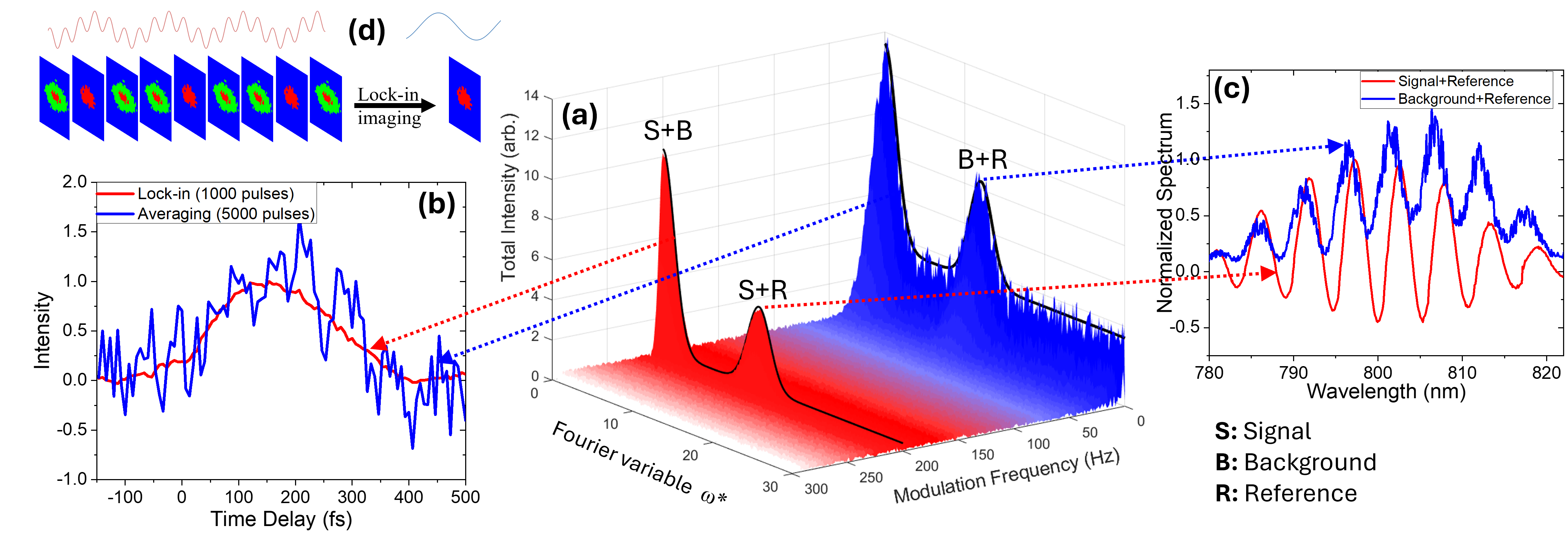}
    \caption{\label{fig:1} Lock-in spectral interferometry scheme showing (a) the total light intensity detected on the spectrometer as a function of the chopper modulation frequency and the Fourier transform variable $\omega^*$ (canonical to optical frequency $\omega$). Incoherent background and other 1/f noise sources form the low-frequency noise seen near 0 Hz modulation frequency. The lineout at 200 Hz represents the $LI,AC$ terms of equation \ref{eq:LIT}. (b) Intensity corresponding to terms without any spectral interference fringes ($\omega^* = 0$) are shown for 0 Hz (blue) and 200 Hz (red) modulation frequency. (c) Intensity of spectral interference ($\omega^* \neq 0$) between the signal and reference (red), and the background and reference (blue) as a function of wavelength. (d) Schematic of lock-in enabled imaging used in the spectrometer.}
\end{figure*}

One solution to remove the coherent background is to use lock-in-enabled detection. The excitation beam(s) can be modulated using an optical chopper, such that the signal of interest also modulates at the chopper frequency, and can be separated from any background that does not modulate. However, any coherent background light falling on the detector will interfere with the signal of interest, giving intensity at the detector

\begin{equation}
    \label{eq:diode}
    I(t_m) = |E_S(t_m) + E_B|^2 = I_S(t_m) + I_B + E_S(t_m) \cdot E_B \, \cos(\phi)
\end{equation}

where terms dependent on $t_m$ are modulated by a chopper. In addition to the signal $I_s$, the interference term here also modulates at the chopper frequency (because $E_S$ is modulated by the optical chopper) and will add to the intensity of the lock-in signal of interest $I_S$. This is an inescapable feature of intensity measurements in the presence of a coherent background. This heterodyning of the signal due to background light is usually ignored, since the background is typically constant. However, such interference can change the relative amplitudes of the real and imaginary parts of signal intensity in a nontrivial manner in addition to adding significant noise due to the presence of the background field. The novel aspect of our implementation lies in the use of a home-built lock-in spectrometer in combination with spectral interferometry that can separate the various interference terms in the measurement. When one of the excitation laser pulse trains is modulated at a given frequency (say 200 Hz) using an optical chopper, the emitted nonlinear signal of interest also modulates at this frequency. All other light falling on the spectrometer does not modulate at the chopper frequency. The terms in equation \ref{eq:RSB} can now be separated into those that modulate (denoted $LI, AC$) and those that do not (denoted $LI, DC$). S, R and B stand for Signal, Reference and Background, respectively:

\begin{equation}
    \begin{split}
        \label{eq:LIT}
        S_{LI,AC} = \ &S_{S}(\omega) + \sqrt{S_{S}(\omega) S_{R}(\omega)}\cos(\varphi_{SR}(\omega)+\omega\, \tau) \\
        &+ \sqrt{S_{S}(\omega) S_{B}(\omega)}\cos(\varphi_{SB}(\omega)) \\ \\
        S_{LI,DC} = \ &S_{R}(\omega) + S_{B}(\omega) +\sqrt{S_{B}(\omega) S_{R}(\omega)}\cos(\varphi_{BR}(\omega) + \omega\, \tau)    
    \end{split}
\end{equation}

Lock-in imaging detection at the chopper frequency can separately measure $S_{LI,AC}$. The interference term of equation \ref{eq:diode} is precisely the last term in the expression of $S_{LI,AC}$ in equation \ref{eq:LIT}. Since this term does not contain an $\omega\tau$ argument in the cosine, it does not lead to any detectable spectral interference fringes and can be filtered out. Thus, this unique combination of lock-in detection and spectral interferometry allows us to isolate the interference of the signal with the reference pulse and reject any contributions from the coherent background. The scheme discussed so far is represented in figure \ref{fig:1}, which shows the 2D conjugate space of lock-in enabled spectral interferometry, with the lock-in frequency on one axis and the frequency of spectral fringes on the other. The three labeled peaks in figure \ref{fig:1} (a) show the three interference terms from equation \ref{eq:RSB}. Figure \ref{fig:1} (b) shows the typical improvement in SNR between lock-in measurement and simple averaging. The separation of interference contributions from different sources enables sampling of multiple regions in the conjugate space to access additional information. One application of this approach, passive stabilization of interference fringes, is discussed in the next section. Analog lock-in amplifiers require separate phase-locked loops (PLLs), mixers, etc. to do simultaneous lock-in measurements at multiple frequencies and thus generally only measure the signal of interest. However, in a digital setting, multiple lock-in measurements can be made simultaneously to sample multiple points in the conjugate space.

To implement lock-in-enabled interferometry, an optical chopper is used to modulate one or more of the excitation beams. The signal light is measured on a CMOS camera, which is triggered at the laser repetition rate. Each camera frame is arranged in a time series (see Figure \ref{fig:1} (d)), where each laser trigger counts as a unit increment of time. Thus, small drifts in the laser pulse arrival time become irrelevant, and the modulation in the signal intensity results only from the chopping, regardless of the pulse repetition rate. The intensity of each pixel in the detector image time series is a 1-dimensional function of time. The chopper trace is acquired in real time using an analog-to-digital converter and fed into a software phase-locked loop (PLL). The chopper trace is a square wave that contains multiple harmonics. Inside the PLL, the acquired chopper trace is Fourier transformed, bandpass filtered around the fundamental frequency, shifted to 0 Hz frequency, and inverse Fourier transformed to retrieve the phase of the fundamental frequency \cite{dixon1989}. From this phase, a sine and a cosine function are constructed and used to perform a lock-in acquisition of the detector pixel intensities at the chopper frequency (dual-phase modulation lock-in amplification). As an aside, it must be noted that since the chopper modulation follows a square wave, the signal modulates as $\sim 1 + \cos(2\pi f t)$ where $f$ is the chopper modulation frequency; thus, a fraction of the signal modulating at the chopper frequency also appears at 0 Hz.

\section{\label{sec:experiments}Application to Experiments}

\begin{figure*}
    \centering
    \includegraphics[width=\linewidth]{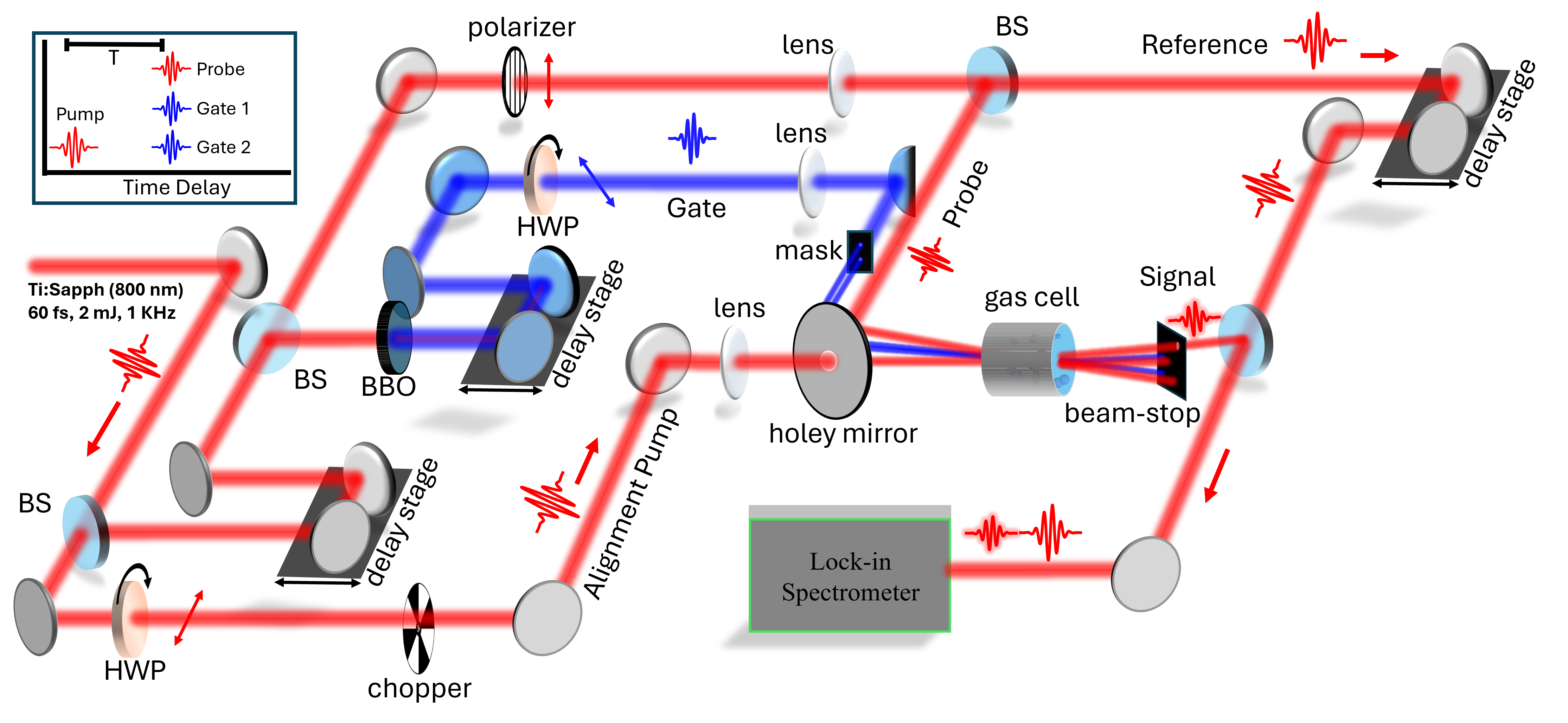}
    \caption{\label{fig:2} Schematic of the experimental setup. The alignment Pump and the time-delayed N-DFWM Probe beams are focused into a gas cell containing the target gas at room temperature and a pressure of 4 bar. The emitted nonlinear signal is spatially isolated, cleaned with a polarizer, and combined with the external reference in a lock-in detection enabled spectrometer. The Reference is separately characterized using a FROG device (not shown). BS: Beam Splitter, HWP: Half-Wave Plate, BBO: Beta-Barium Borate (nonlinear crystal).}
\end{figure*}

In contrast to single-/double-channel lock-in spectrometers used with MHz repetition rate lasers \cite{emde1997,uhl2021}, the KHz repetition rate of high-powered lasers typically used in nonlinear optical experiments, such as high-harmonic spectroscopy, makes real-time multichannel digital lock-in analysis feasible. To demonstrate lock-in enabled interferometry, we describe a non-degenerate four-wave mixing (N-DFWM) experiment in gas phase molecules. In the experiment, 60-fs near-infrared (IR) pulses centered at 800 nm are split once and delayed. One arm forms the Alignment Pump beam, while the other is split again to create the N-DFWM beams (see Figure \ref{fig:2}). The three N-DFWM beams are labeled Gate 1, Gate 2 and Probe. A single Gate beam is frequency doubled using a beta-barium borate (BBO) crystal and delayed with respect to the probe beam. The Gate beam is then sent through a spatial mask to derive the co-timed Gate 1 and Gate 2 beams. Further, a fraction of the Probe is picked off and delayed, to form the external Reference beam. The time delay between the Alignment Pump and the Probe is denoted by T, while the time delay between the two Gates and the Probe is denoted by $\tau$. The polarization of the Alignment Pump is set orthogonal to the Probe using a half-wave plate (HWP). The polarization of the Gate pulses can be rotated to measure different components of the nonlinear response tensor. All beams are focused into a gas cell with room-temperature nitrogen gas at a pressure of 4 bar. The gas cell is $\sim 90$ mm long, with 1 mm thick UV fused silica (UVFS) windows. The time delay between the Alignment Pump and N-DFWM pulses (T) is scanned, while the time delay between the Gate and Probe pulses ($\tau$) is set to zero. In this folded BOXCARS configuration \cite{shirley1980}, the emitted signal travels in a different direction than the other beams due to phase matching. The Signal and Reference beams are sent into the home-built lock-in spectrometer. An optical chopper is placed in the path of the Alignment Pump. 

\begin{figure*}
    \centering
    \includegraphics[width=\linewidth]{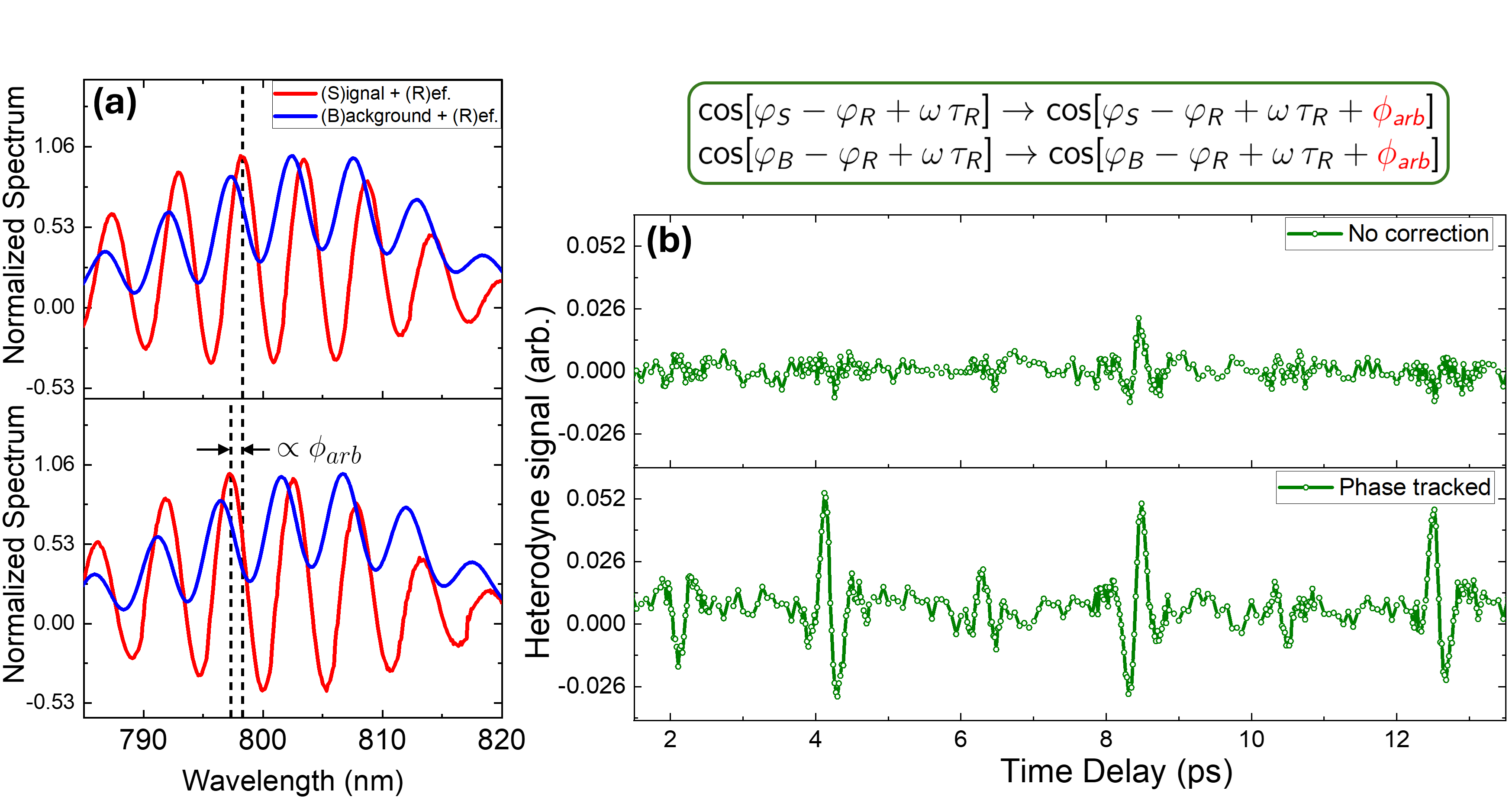}
    \caption{\label{fig:3} (a) Interference fringes between the signal and reference (red), and the background and reference (blue) are shown for two consecutive measurements. Even though the fringes drift between the two measurements, they are phase-locked relative to each other. (b) Experimental heterodyne signal corresponding to the frequency integrated Signal-Reference interference term in equation \ref{eq:LIT} with and without phase-tracking. Without any drift correction from phase tracking, the measured heterodyne signal phase drifts between each measurement and vanishes when averaged.}
\end{figure*}

\begin{figure}
    \centering
    \includegraphics[width=0.5\linewidth]{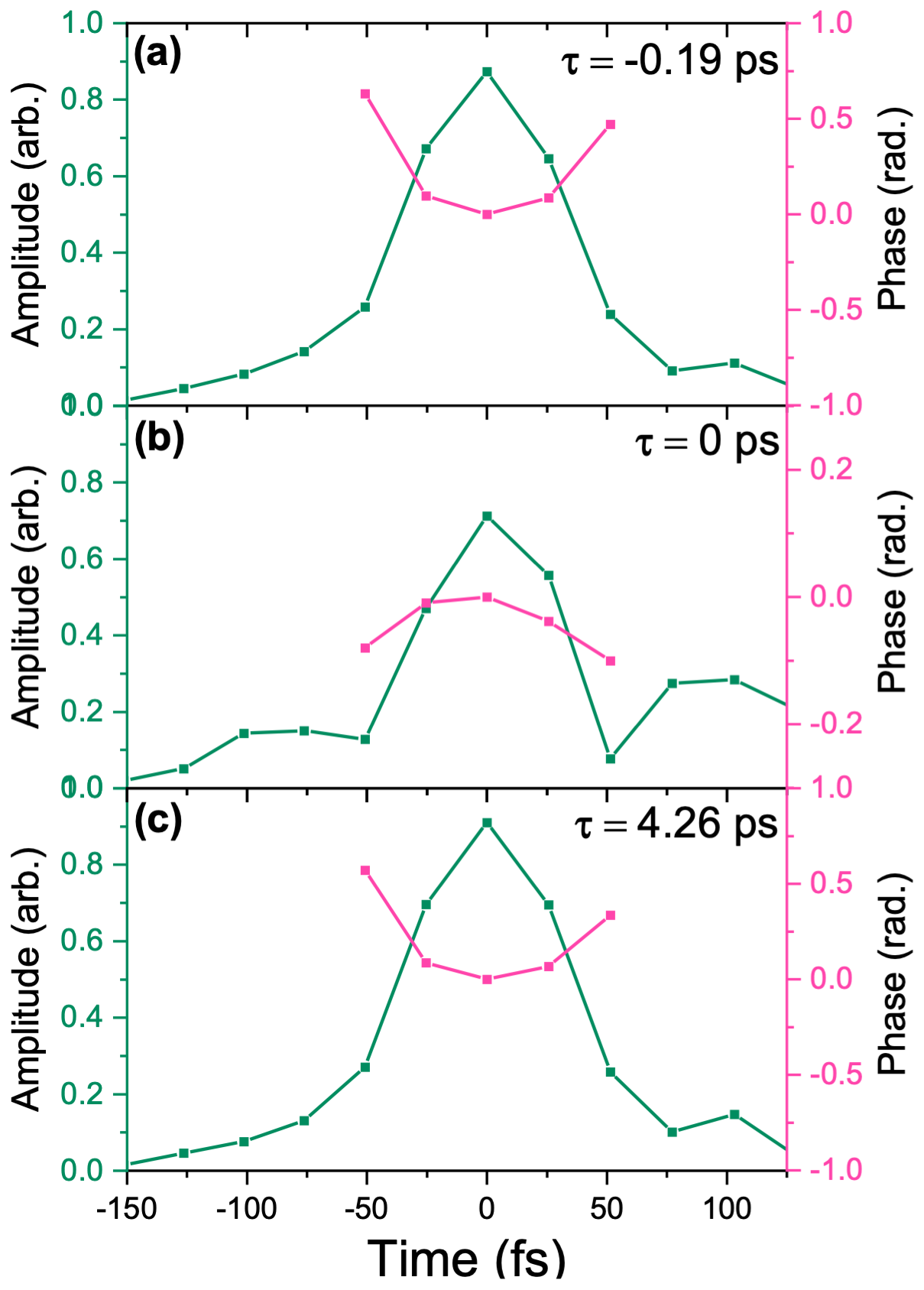}
    \caption{\label{fig:4} Measured signal field (temporal amplitude and phase) at three different pump-probe time delays. (a) At negative time delay, the pulse is positively chirped. (b) At pump-probe overlap, the intense pump pulse causes ionization, leading to a negatively chirped signal pulse. (c) At $\tau = 4.26$ ps, when the molecule is anti-aligned, the chirp is lowered.}
\end{figure}

Drifts in the optical path length between the Reference and the Signal arms of the interferometer lead to spectral fringe instability. Such drifts can be caused by temperature changes, vibrations, or air currents. We can model such drifts by adding an arbitrary phase offset $\phi_{arb}$ to all interference terms that contain $\omega \tau$. Although faster drifts (jitter) are efficiently removed by lock-in amplification, slower drifts remain. Usually, such slower drifts are corrected by sending a "tracer beam" along the path of the interferometer to track phase drifts between the interferometer arms in real time \cite{gallagher1998,volkov2005,belabas2002,zhang2005}. 
Interferometric stability between excitation pulses has also been achieved by using lock-in detection, without introducing an additional tracer beam \cite{tekavec2006,bruder2015}. However, this technique is not applicable when one of the pulses acts as a pump, while the other acts as probe. Lock-in-enabled spectral interferometry, as demonstrated here, provides an alternative approach to remove phase drifts. Since the signal and background (from unaligned molecules) travel along the same path, any environment-dependent phase drifts will show up in both sets of fringes, corresponding to Signal-Reference interference and Signal-Background interference, equally. For each delay step, the acquired spectrometer image can be lock-in filtered at both the chopper frequency and 0 Hz, and $\phi_{arb}$ can be removed by subtracting the phase of the background fringes from that of the signal fringes, without needing active stabilization. Without this phase tracking, the measured N-DFWM heterodyne signal averages to zero (see Figure \ref{fig:3}). On the other hand, setting the spectral phase of each measurement to zero at the central wavelength leads to a non-vanishing signal but it contains no phase-shift information acquired during the nonlinear interaction. To accurately measure the real and imaginary parts of the signal E-field, phase tracking is necessary. In figure \ref{fig:4}, we show the measured signal E-field temporal amplitude and phase, obtained by applying a fourier transform to the spectral domain electric field, for three different pump-probe time delays. For negative pump-probe time delays, the phase is positively chirped, while around a rotational revival the chirp modulates \cite{comstock2003,bartels2001}. At zero time delay, the negative dispersion caused by free electrons created by the pump leads to a negatively chirped signal pulse \cite{hauri2006}. Although the temporal shape of the phase of the signal electric field can be measured without phase tracking and contains useful information \cite{walz2022, pandey2024}, the net phase shift of the signal is necessary in many measurements. Lock-in-enabled phase tracking presented here can be a powerful approach to measure such phase shifts, especially in the presence of a coherent background, without the need for active stabilization. 

\section{\label{sec:conclusion}Conclusions}
Measuring the full E-field in nonlinear optical experiments, as a function of time delay between the interacting pulses, can provide detailed insight into the dynamics of excited molecules. In this work, we have demonstrated that lock-in detection enabled spectral interferometry can resolve very weak signals (pJ - fJ) in nonlinear spectroscopy experiments, even in the presence of a large coherent background. This makes lock-in-enabled spectral interferometry a strong candidate for ultrafast all-optical experiments, including those that use high-harmonic sources for pumping molecules. The low flux of high-harmonic sources places an upper limit on the number of signal photons generated, making the use of many other direct field sampling methods \cite{liu2021,liu2022,bionta2021,park2018,hui2022,schiffrin2013,paasch2014,paasch2016,sederberg2020} that require higher pulse energies challenging. Our work also highlights the benefits of the multidimensionality of the conjugate space of lock-in enabled spectral interferometry. By measuring the interference fringes at two different lock-in frequencies, we have demonstrated a method for stabilizing interference fringes by tracking changes in the optical path. To reject any scattered light from the chopper-modulated laser beam, a second chopper could be used in the path of a different laser beam, and lock-in detection could be performed at the difference/sum frequency. The conjugate space of such an experiment will look more complex and might provide new options for the correction of experimental imperfections, similar to the phase tracking demonstrated in this work.

\begin{backmatter}
\bmsection{Funding}
 Content in the funding section will be generated entirely from details submitted to Prism. Authors may add placeholder text in the manuscript to assess length, but any text added to this section in the manuscript will be replaced during production and will display official funder names along with any grant numbers provided. If additional details about a funder are required, they may be added to the Acknowledgments, even if this duplicates information in the funding section. See the example below in Acknowledgements.

\bmsection{Acknowledgments}
This material is based upon work supported by the National Science Foundation under Grant No. 2208061.

\bmsection{Disclosures}
The authors declare no conflicts of interest.

\bmsection{Data availability} Data underlying the results presented in this paper are not publicly available at this time but may be obtained from the authors upon reasonable request.


\end{backmatter}

\bibliography{references}

\end{document}